# Points on Computable Curves
## (Extended Abstract)

Xiaoyang Gu [*][‡]    Jack H. Lutz [†][‡]    Elvira Mayordomo [§][¶]

The "analyst's traveling salesman theorem" of geometric measure theory characterizes those subsets of Euclidean space that are contained in curves of finite length. This result, proven for the plane by Jones (1990) and extended to higher-dimensional Euclidean spaces by Okikiolu (1991), says that a bounded set $K$ is contained in some curve of finite length if and only if a certain "square beta sum", involving the "width of $K$" in each element of an infinite system of overlapping "tiles" of descending size, is finite.

In this paper we characterize those *points* of Euclidean space that lie on *computable* curves of finite length. We do this by formulating and proving a computable extension of the analyst's traveling salesman theorem. Our extension, the *computable analyst's traveling salesman theorem*, says that a point in Euclidean space lies on some computable curve of finite length if and only if it is "permitted" by some computable "Jones constriction". A Jones constriction here is an explicit assignment of a rational cylinder to each of the above-mentioned tiles in such a way that, when the radius of the cylinder corresponding to a tile is used in place of the "width of $K$" in each tile, the square beta sum is finite. A point is permitted by a Jones constriction if it is contained in the cylinder assigned to each tile containing the point. The main part of our proof is the construction of a computable curve of finite length traversing all the points permitted by a given Jones constriction. Our construction uses the main ideas of Jones's "farthest insertion" construction, but our algorithm for computing the curve must work exclusively with the Jones constriction itself, because it has no direct access to the (typically uncomputable) points permitted by the Jones constriction.

---

[*]Department of Computer Science, Iowa State University, Ames, IA 50011, USA. Email: xiaoyang@cs.iastate.edu

[†]Department of Computer Science, Iowa State University, Ames, IA 50011, USA. Email: lutz@cs.iastate.edu

[‡]Research supported in part by National Science Foundation Grant 0344187.

[§]Departamento de Informática e Ingeniería de Sistemas, Universidad de Zaragoza, 50018 Zaragoza, Spain. Email: elvira@unizar.es

[¶]Research supported in part by Spanish Government MEC Project TIC 2002-04019-C03-03.

# 1 Introduction

Where can an infinitely small robot go? This paper answers a precise form of this fanciful question by formulating and proving a computable extension of the celebrated "analyst's traveling salesman theorem" of geometric measure theory.

The precise statement of our question is straightforward. Our robot is the size of a geometric point (the "ultimate nanobot"), and it moves in a Euclidean space $\mathbb{R}^n$, where $n \geq 2$. The robot's motion is algorithmic, and there are no obstacles, thermal effects, or quantum effects, so its path is a computable curve, i.e., a curve traced by a computable function $f : [0, 1] \to \mathbb{R}^n$. The robot's path has arbitrary but finite length. (The computable curve is *rectifiable*. Among other things, this implies that it is not a space-filling curve [20].) The robot's motion is otherwise unrestricted. For example, it may cross or retrace its own path, so the function $f$ is not required to be one-to-one. (In the terminology of some, $f$ describes a tour, rather than a curve. In the terminology of others, $f$ describes a curve that need not be simple.)

The collection of all possible paths of our robot forms a "computable transit network" RCC $\subseteq \mathbb{R}^n$. This is the union of all rectifiable computable curves in $\mathbb{R}^n$, i.e., the set of all points $x \in \mathbb{R}^n$ such that some computable curve of finite length passes through $x$. Our question is simple. Which points in $\mathbb{R}^n$ lie in the set RCC?

A brief summary of some basic properties of RCC (developed in detail in section 3) sets the stage for our main results. It is easy to see that RCC has Hausdorff dimension 1, so most points in $\mathbb{R}^n$ do not lie in RCC. On the other hand, RCC is a dense subset of $\mathbb{R}^n$, and RCC is path-connected in the strong sense that any two points in RCC lie on a *single* computable curve of finite length. Each point $x \in$ RCC has dimension at most 1 (by which we mean that $\{x\}$ has constructive dimension at most 1 [13]), but the complement of RCC contains points of arbitrarily small dimension, so this does not characterize membership in RCC.

Our main theorem characterizes points in RCC by extending the famous "analyst's traveling salesman theorem" of geometric measure theory to a theorem in computable analysis. The analyst's traveling salesman theorem, proven for $\mathbb{R}^2$ by Jones in 1990 [8] and extended to $\mathbb{R}^n$ for $n \geq 2$ by Okikiolu in 1991 [18] (see also the monographs [15, 5]), gives a precise characterization of those subsets of $\mathbb{R}^n$ that are contained in rectifiable curves.

For each $m \in \mathbb{Z}$, let $\mathcal{Q}_m$ be the set of all *dyadic cubes of order* $m$, which are half-closed, half-open cubes
$$Q = [a_1, a_1 + 2^{-m}) \times \cdots \times [a_n, a_n + 2^{-m})$$
in $\mathbb{R}^n$ with $a_1, \ldots, a_n \in 2^{-m}\mathbb{Z}$. Note that such a cube $Q$ has sidelength $\ell(Q) = 2^{-m}$ and all its vertices in $2^{-m}\mathbb{Z}^n$. Let $\mathcal{Q} = \bigcup_{m \in \mathbb{Z}} \mathcal{Q}_m$ be the set of all dyadic cubes of all orders. We regard each dyadic cube $Q$ as an "address" of the larger cube $3Q$, which has the same center as $Q$ and sidelength $\ell(3Q) = 3\ell(Q)$. The analyst's traveling salesman theorem is stated in terms of the resulting system $\{3Q \mid Q \in \mathcal{Q}\}$ of overlapping cubes.

Let $K$ be a bounded subset of $\mathbb{R}^n$. For each $Q \in \mathcal{Q}$, let $r(Q)$ be the least radius of any infinite closed cylinder in any direction in $\mathbb{R}^n$ that contains all of $K \cap 3Q$. Then the *Jones* *beta-number* of $K$ at $Q$ is

$$\beta_Q(K) = \frac{r(Q)}{\ell(Q)},$$

and the *Jones square beta-number* of $K$ is

$$\beta^2(K) = \sum_{Q \in \mathcal{Q}} \beta_Q(K)^2 \ell(Q)$$

(which may be infinite). Here is the analyst's traveling salesman theorem.

**Theorem 1.1.** *(Jones [8], Okikiolu [18]). Let $K \subseteq \mathbb{R}^n$ be bounded. Then $K$ is contained in some rectifiable curve if and only if $\beta^2(K) < \infty$.*

Jones's proof of the "if" direction of Theorem 1.1 is an intricate "farthest insertion" construction of a curve containing $K$, together with an amortized analysis showing that the length of this curve is finite. This proof works in any Euclidean space $\mathbb{R}^n$. However, Jones's proof of the "only if" direction of Theorem 1.1 uses nontrivial methods from complex analysis and only works in the Euclidean plane $\mathbb{R}^2$ (regarded as the complex plane $\mathbb{C}$). Okikiolu's subsequent proof of the "only if" direction is a clever geometric argument that works in any Euclidean space $\mathbb{R}^n$. (It should also be noted that these papers establish a quantitative relationship between $\beta^2(K)$ and the infimum length of a curve containing $K$, and that the constants in this relationship have been improved in the recent thesis by Schul [21].)

Theorem 1.1 is generally regarded as a solution of the "analyst's traveling salesman problem" (analyst's TSP), which is to characterize those sets $K \subseteq \mathbb{R}^n$ that can be traversed by curves of finite length. It is then natural to pose the *computable analyst's TSP*, which is to characterize those sets $K \subseteq \mathbb{R}^n$ that can be traversed by *computable* curves of finite length. While the analyst's TSP is only interesting for infinite sets $K$ (because *every* finite set $K$ is contained in a rectifiable curve), the computable analyst's TSP is interesting for arbitrary sets $K$. In fact, the question posed at the beginning of this introduction is precisely the computable analyst's TSP restricted to singleton sets $K = \{x\}$.

To solve the computable analyst's TSP, we first have to replace the Jones square beta-number of the arbitrary set $K$ with a data structure that can be required to be computable. To this end, we define a *cylinder assignment* to be a function $\gamma$ assigning to each dyadic cube $Q$ an (infinite) closed *rational* cylinder $\gamma(Q)$, by which we mean that $\gamma(Q)$ is a cylinder whose axis passes through two (hence infinitely many) points of $\mathbb{Q}^n$ and whose radius $\rho(Q)$ is rational. (If $\rho(Q) = 0$, the cylinder is a line; if $\rho(Q) < 0$, the cylinder is empty.) The *set permitted by* a cylinder assignment $\gamma$ is the (closed) set $\kappa(\gamma)$ consisting of all points $x \in \mathbb{R}^n$ such that, for all $Q \in \mathcal{Q}$,

$$x \in (3Q)^o \Rightarrow x \in \gamma(Q),$$

where $(3Q)^o$ is the interior of $3Q$.

There is one technical point that needs to be addressed here. If $\gamma$ is a cylinder assignment that, at some $Q \in \mathcal{Q}$, prohibits a subcube $3Q'$ of $3Q$ (i.e., $\gamma(Q) \cap (3Q')^o = \varnothing$), then $\kappa(\gamma)$ contains no interior point of $3Q'$, so it is pointless and misleading for $\gamma$ to assign $Q'$ a cylinder $\gamma(Q')$ that meets $(3Q')^o$. We define a cylinder assignment $\gamma$ to be *persistent* if it does not make such pointless assignments, i.e., if, for all $Q, Q' \in \mathcal{Q}$ with $Q' \subseteq Q$ and $\gamma(Q) \cap (3Q')^o = \emptyset$, we have $\gamma(Q') \cap (3Q')^o = \emptyset$. It is easy to transform a cylinder assignment $\gamma$ into a persistent cylinder assignment $\gamma'$ that is equivalent to $\gamma$ in the sense that $\kappa(\gamma) = \kappa(\gamma')$, with $\gamma'$ computable if $\gamma$ is.

**Definition.** Let $\gamma$ be a cylinder assignment.

1. The *Jones beta-number* of $\gamma$ at a cube $Q \in \mathcal{Q}$ is
$$\beta_Q(\gamma) = \frac{\rho(Q)}{\ell(Q)}.$$

2. The *Jones square beta-number* of $\gamma$ is
$$\beta^2(\gamma) = \sum_{Q \in \mathcal{Q}} \beta_Q(\gamma)^2 \ell(Q).$$

Note that $\beta^2(\gamma)$ may be infinite.

**Definition.** A *Jones constriction* is a persistent cylinder assignment $\gamma$ for which $\beta^2(\gamma) < \infty$.

We can now state our main result, the computable analyst's traveling salesman theorem.

**Theorem 1.2.** *Let $K \subseteq \mathbb{R}^n$ be bounded. Then $K$ is contained in some rectifiable computable curve if and only if there is a computable Jones constriction $\gamma$ such that $K \subseteq \kappa(\gamma)$.*

Theorem 1.2 solves the computable analyst's TSP, and thus immediately solves our question about where an infinitely small robot can go:

**Corollary 1.3.** *A point $x \in \mathbb{R}^n$ lies on some computable curve of finite length if and only if $x$ is permitted by some computable Jones constriction. That is,*
$$\mathrm{RCC} = \bigcup_{\text{computable } \gamma} \kappa(\gamma),$$
*where the union is taken over all computable Jones constrictions.*

It should be noted that (the proof of) Theorem 1.2 relativizes to arbitrary oracles, so it implies Theorem 1.1. This is the sense in which our computable analyst's traveling salesman theorem is an extension of the analyst's traveling salesman theorem.

Our proof of the "only if" direction of Theorem 1.2 is easy, because we are able to use the corresponding part of Theorem 1.1 as a "black box". However, our proof of the "if" direction is somewhat involved. Given an arbitrary computable Jones constriction $\gamma$, we construct a rectifiable computable curve containing $\kappa(\gamma)$. In this construction, we are able to follow the broad outlines of Jones's "farthest insertion" construction and to use its key ideas, but we have an additional obstacle to overcome. The analyst's TSP does not require an algorithm, so Jones's proof can simply "choose" elements of the given set $K$ according to various criteria at each stage of the construction (often moving these points later as needed). However, even if $\gamma$ is computable, neither the set $\kappa(\gamma)$ nor its elements need be computable. Hence the algorithm for our computable curve cannot directly choose points in (or even reliably near) $\kappa(\gamma)$. Our construction succeeds by carefully separating the algorithm from the amortized analysis of the length of the curve that it computes. The proof is discussed in some detail in section 4 and at greater length in the appendix.

## 2 Curves and Computability

We fix an integer $n \geq 2$ and work in the Euclidean space $\mathbb{R}^n$. A *curve* is a continuous function $f : [0,1] \to \mathbb{R}^n$. The *length* of a curve $f$ is

$$\text{length}(f) = \sup_{\vec{a}} \sum_{i=0}^{k-1} |f(a_{i+1}) - f(a_i)|,$$

where $|x|$ is the Euclidean norm of a point $x \in \mathbb{R}^n$ and the supremum is taken over all *dissections* $\vec{a}$ of $[0,1]$, i.e., all $\vec{a} = (a_0, \ldots, a_k)$ with $0 = a_0 < a_1 < \cdots < a_k = 1$. Note that length($f$) is the length of the actual path traced by $f$. If $f$ is one-to-one (i.e., the curve is *simple*), then length($f$) coincides with $\mathcal{H}^1(f([0,1]))$, which is the length (i.e., the one-dimensional Hausdorff measure [4]) of the range of $f$, but, in general, $f$ may "retrace" parts of its range, so length($f$) may exceed $\mathcal{H}^1(f([0,1]))$. A curve $f$ is *rectifiable* if length($f$) $< \infty$.

A *tour* of a set $K \subseteq \mathbb{R}^n$ is a curve $f : [0,1] \to \mathbb{R}^n$ such that $K \subseteq f([0,1])$.

Since curves are continuous, the extended computability notion introduced by Braverman [1] coincides with the computability notion formulated in the 1950s by Grzegorczyk [6] and Lacombe [10] and exposited in the recent paper by Braverman and Cook [2] and in the monographs [19, 9, 23]. Specifically, a curve $f : [0,1] \to \mathbb{R}^n$ is *computable* if there is an oracle Turing machine $M$ with the following property. For all $t \in [0,1]$ and $r \in \mathbb{N}$, if $M$ is given a function oracle $\varphi_t : \mathbb{N} \to \mathbb{Q}$ such that, for all $k \in \mathbb{N}$, $|\varphi_t(k) - t| \leq 2^{-k}$, then $M$, with oracle $\varphi_t$ and input $r$, outputs a rational point $M^{\varphi_t}(r) \in \mathbb{Q}^n$ such that $|M^{\varphi_t}(r) - f(t)| \leq 2^{-r}$.

A point $x \in \mathbb{R}^n$ is *computable* if there is a computable function $\psi_x : \mathbb{N} \to \mathbb{Q}^n$ such that, for all $r \in \mathbb{N}$, $|\psi_x(r) - x| \leq 2^{-r}$. It is well known and easy to see that, if $f : [0,1] \to \mathbb{R}^n$ and $t \in [0,1]$ are computable, then $f(t)$ is computable.

## 3 The Set RCC

As in the introduction, we let RCC denote the set of all points in $\mathbb{R}^n$ that lie on rectifiable computable curves. We briefly discuss the structure of RCC, referring freely to existing literature on fractal geometry [4] and constructive dimension [13].

For each rectifiable curve $f$, we have $\mathcal{H}^1(f([0,1])) \leq \text{length}(f) < \infty$, so the Hausdorff dimension of $f([0,1])$ is 1, unless $f([0,1])$ is a single point (in which case the Hausdorff dimension is 0). Since RCC is the union of countably many such sets $f([0,1])$, it follows by countable stability that RCC has Hausdorff dimension 1 [4]. This implies that RCC is a Lebesgue measure 0 subset of $\mathbb{R}^n$, i.e., that almost every point in $\mathbb{R}^n$ lies in the complement of RCC.

Since RCC contains every computable point in $\mathbb{R}^n$, RCC is dense in $\mathbb{R}^n$. Also, if $x \in f([0,1])$ and $y \in g([0,1])$, where $f$ and $g$ are rectifiable computable curves, then we can use $f$, $g$, and the segment from $f(1)$ to $g(0)$ to assemble a rectifiable computable curve $h$ such that $x, y \in h([0,1])$. Hence, RCC is path-connected in the strong sense that any two points in RCC lie in a *single* rectifiable computable curve.

For each rectifiable computable curve $f$, the set $f([0,1])$ is a computably closed (i.e., $\Pi_1^0$) subset of $\mathbb{R}^n$ [17]. Since RCC is the union of all such $f([0,1])$, it follows by Hitchcock's correspondence principle [7] that the constructive dimension of RCC coincides with its Hausdorff dimension, which we have observed to be 1. (It is worth mention here that RCC can easily be shown *not* to have computable measure 0, whence RCC has computable dimension $n$ [12]. By Staiger's correspondence principle [22], this implies that RCC is not a $\Sigma_2^0$ set.) It follows that each point $x \in$ RCC has dimension at most 1 (in the sense that $\{x\}$ has constructive dimension 1 [13]). It might be reasonable to conjecture that this actually characterizes points in RCC, but the following example shows that this is not the case.

**Example 3.1.** Given an infinite binary sequence $R$, define a sequence $A_0, A_1, A_2, \ldots$ of closed squares in $\mathbb{R}^2$ by the following recursion. First, $A_0 = [0,1]^2$. Next, assuming that $A_n$ has been defined, let $a$ and $b$ be the $2n$th and $(2n+1)$st bits, respectively of $R$. Then $A_{n+1}$ is the $ab$-most closed subsquare of $A_n$ with $\text{area}(A_{n+1}) = \frac{1}{16}\text{area}(A_n)$, where $00 =$ "lower left", $01 =$ "lower right", $10 =$ "upper left", and $11 =$ "upper right". Let $x_R$ be the unique point in $\mathbb{R}^2$ such that $x_R \in A_n$ for all $n \in \mathbb{N}$. It is well known [16, 5] that the set $K$, consisting of all such points $x_R$, is a bounded set with positive, finite one-dimensional Hausdorff measure (and hence with Hausdorff dimension 1), but that $K$ is not contained in any rectifiable curve. A constructive extension of this proof shows that, for any sequence $R$ that is random (in the sense of Martin-Löf[14]; see also [11, 3]), the point $x_R$ has dimension 1 and does not lie on any computable curve of finite length.

The following theorem shows that more is true, although the proof, a Baire category argument, does not yield such a concrete example.

**Theorem 3.2.** *The complement of* RCC *contains points of arbitrarily small dimension, including* 0.

# 4 The Computable Analyst's Traveling Salesman Theorem

This section presents the main ideas of the proof of Theorem 1.2. The detailed proof appears in preliminary form in the appendix.

We first dispose of the "only if" direction. If we are given a rectifiable computable curve $f$ and a rational $\epsilon > 0$, it is routine to construct a computable Jones constriction $\gamma$ such that $f([0,1]) \subseteq \kappa(\gamma)$ and $\beta^2(\gamma) \leq \beta^2(f([0,1])) + \epsilon$. The "only if" direction of Theorem 1.2 hence follows easily from the "only if" direction of Theorem 1.1. We thus focus our attention on proving the "if" direction of Theorem 1.2.

As pointed out by Jones [8], the analyst's TSP is significantly different from the classical TSP in that it typically involves uncountably many points at locations that are not explicitly specified. In his construction, he has the privilege to "know" whether a point is in the set $K$ or not, since he is concerned only with the existence of a tour and not with the computability of the tour. This is no longer true in our situation, since we work with only a computable constriction, from which we may not computably determine whether a point is in the set. Although the situations differ by so much, ideas with a flavor of the "farthest insertion" and "nearest insertion" heuristics that are used in Jones's argument and the classical TSP are essential parts of our solution.

Given a computable Jones constriction $\gamma$, we construct computably a tour $f : [0,1] \to \mathbb{R}^n$ of the set $K = \kappa(\gamma)$ permitted by $\gamma$ such that $\kappa(\gamma) \subseteq f([0,1])$ and the length of the tour is finite.

Our construction proceeds in stages. In each stage $m \in \mathbb{N}$, a set of points with regulated density is chosen according to the constriction and a tour $f_m$ of these points is constructed so that every point in $K$ is at most roughly $2^{-m}$ from the tour. Every tour is constructed by patching the previous tour locally so that the sequence of tours $\{f_m\}$ converges computably.

During the tour patching at each stage, the insertion ideas mentioned earlier are applied at different parts of the set $K$ according to the local topology given by the constriction. Note that it is not completely clear that the use of "farthest insertion" is absolutely necessary. However, it greatly facilitates the associated amortized analysis of length, which is as crucial in our proof as it is in Jones's. In the following, we describe in more detail how and when these ideas are applied in the algorithmic construction of the tour.

In each stage $m \in \mathbb{N}$, we look at cubes $Q$ of sidelength $A2^{-m}$, where $A = 2^{k_0}$ is a sufficiently large universal constant. We pick points so that they are at least $2^{-m}$ from each other and every point in $K$ is at most $2^{-m}$ from some of those chosen points. Based on the value of $\beta_Q(\gamma)$, which measures the relative width of $3Q \cap K$, we divide cubes into "narrow" ones ($\beta_Q(\gamma) < \epsilon_0$) and "fat" ones ($\beta_Q(\gamma) \geq \epsilon_0$), where $\epsilon_0$ is a small universal constant.

The fat cubes are easy to process, since the associated square beta-number is large. We connect the points in those cubes to nearby surrounding points, some of which are guaranteed to be in the previous tour due to the density of the points in the tour. Since the points are chosen with regulated density, the number of connections we make here is bounded by a universal constant. The length of each connection is proportional to the sidelength of the cube, which is proportional to $2^{-m}$. Thus the total length we add to the tour is bounded by $c_0 \cdot \epsilon_0^2 \ell(Q)$, which is then bounded by $c_0 \cdot \beta_Q^2(\gamma) \ell(Q)$, where $c_0$ is a sufficiently large universal constant.

For the narrow cubes, we carry out either "farthest insertion" or "nearest insertion" depending on the local topology around each insertion point.

Suppose that we are about to patch the existing tour to include a point $x$. Since from stage to stage, the points are picked with increasing density, there is always a point $z_1$ already in the tour inside the cube that contains $x$. However, there are two possibilities for the neighborhood of $x$. One is that there is another point $z_2$ already in the tour and $z_2$ is inside the cube that contains $x$. The other possibility is that $z_1$ is the only such point.

In the first case, point $x$ lies in a narrow cube and there are points $z_1$ and $z_2$ in the narrow cube such that $x$ is between $z_1$ and $z_2$. Points $z_1$ and $z_2$ are in the existing tour and are connected directly with a line segment in the tour. In this case, we apply "nearest insertion" by letting $z_1$ and $z_2$ be the closest two neighbors of $x$ in the existing tour, breaking the line segment between $z_1$, $z_2$, and connecting $z_1$ to $x$ and $x$ to $z_2$. The increment of the length of the tour is $\ell([z_1, x]) + \ell([x, x_2]) - \ell([z_1, z_2])$, which is bounded by $c_1 \cdot \beta_Q^2(\gamma)\ell(Q)$ by an application of the Pythagorean theorem, since the cube is very narrow.

In the second case, point $z_1$ is the only point in the existing tour that is in the same cube as $x$. It is not guaranteed that $x$ can be inserted between two points in the existing tour. Even when it is possible, the other point in the existing tour would be outside the cube that we are looking at and thus it might require backtracking an unbounded number of stages to bound the increment of length, which would make the proof extremely complicated (if even possible). Therefore, we keep the patching for every point local and, in this case, we make sure $x$ is locally the "farthest" point from $z_1$ and connect $x$ directly to $z_1$. (Note that the actual situation is slightly more involved and is addressed in the full proof.) In this case, the Pythagorean theorem cannot be used and thus we cannot use the Jones square beta-number to directly bound the increment of length. To remedy this, we employ amortized analysis and save spare square beta-numbers in a savings account over the stages and use the saved values to bound the length increment. In order for this to work, we choose $\epsilon_0$ so small that at a particular neighborhood, "farthest insertion" does not happen very frequently and we always have the time to save up enough of the square beta-number before we need to use it.

**Acknowledgment.** The second author thanks Dan Mauldin for pointing out the existence of [8].8

# A  Technical Appendix

## A.1  Pythagorean Theorem

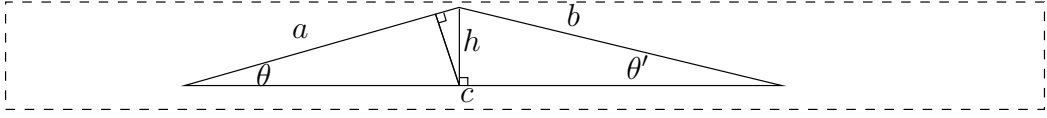

Figure A.1: Pythagorean Theorem

**Theorem A.1.** *Let $m \in \mathbb{Z}$ and $A > 9$. Let $a, b, c$ be the lengths of three line segments that form a triangle inside a cylinder of length $l = A2^{1-m}$ and width $w < \frac{l}{A^3 \sqrt{n}}$ such that $2^{1-m} \geq a, b \geq 2^{-m}$ and $c \geq 2^{1-m}$, where $n$ is dimension of the space. Let $\beta = \frac{w}{l}$. Then*

$$a + b \leq c + 2A\beta^2 l.$$

*Proof.* Let $\theta$ be the small angle determined by line segments $a$ and $c$. Let $\theta'$ be the small angle determined by line segments $b$ and $c$. Let $h$ be the distance from the intersection of line segments $a$ and $b$ to line segment $c$.

$$\begin{aligned}
a + b - c &\leq h \sin\theta + h \sin\theta' = h \cdot \frac{h}{a} + h \cdot \frac{h}{b} \\
&= a \cdot \left(\frac{h}{a}\right)^2 + b \cdot \left(\frac{h}{b}\right)^2 \leq 2A \left(\frac{w}{l}\right)^2 \cdot l \\
&= 2A\beta^2 l.
\end{aligned}$$

□

This version of Pythagorean Theorem easily generalizes to the case where more line segments are involved in the setting.

## A.2  The Construction

Note that by the definition of constriction, the set $K = \kappa(\gamma)$ permitted by constriction $\gamma$ is compact. Without loss of generality, we assume $K \subseteq [0, 1/\sqrt{n}]^n$, $(0, \ldots, 0) \in K$, and $(1/\sqrt{n}, \ldots, 1/\sqrt{n}) \in K$. Let $A = 2^{k_0} > 9$. Let $\epsilon_0 < \frac{1}{A^3 \sqrt{n}}$ be a fixed small constant, where $n$ is the dimension of the Euclidean space we are working with.

In the construction, we inductively build point sets $L_0 \subseteq L_1 \subseteq \cdots \subseteq L_m \cdots$ in stages with the following properties.

C1: $|z_j - z_k| \geq 2^{-m} - \sqrt{m} 2^{-2^m}$, for $z_j, z_k \in L_m$, $j \neq k$.

C2: For $m \in \mathbb{N}$ and every $x \in K$, there exists $z \in L_m$ such that $|x - z| \leq 2^{-m} + \sqrt{m} 2^{-2^m}$.

Note that for each $m \in \mathbb{N}$, $L_m \subseteq K_m$, where $K_m$ is the union of dyadic cubes of sidelength $2^{-2^m}$ permitted by $\gamma$. However, the points in $L_m$ are not specified by explicit coordinates. Instead, every point in $L_m$ is specified by an algorithm, which when given a precision parameter $r$, outputs the coordinates of the dyadic cube of sidelength at most $2^{-r}$ that the point lies in. At stage $m$, we use $r = 2^m$. Although the points we pick may not have rational coordinates, at each stage $m$, we only look at them with precision $r$ and treat them as if they all have rational coordinates. The dyadic cube determined by the coordinates is a sub-cube of the dyadic cube given by smaller precision parameter $m$. Thus the point is specified by a nested chain of dyadic cubes of progressively smaller sizes. When, for some $m$, such a dyadic cube is not permitted by $\gamma$, the output of the algorithm remains to be coordinates given by the algorithm with the largest precision parameter that leads to an output of a dyadic cube that is permitted by $\gamma$. Thus it is possible that a point in $L_m$ is not in $K$.

In stage $m \in \mathbb{N}$, we look at cubes $Q$ of sidelength $A2^{-m}$. For each $Q$, we use $3Q$ to denote the cube of side length $3A2^{-m}$ centered at the center of $Q$. For the sake of precision, we look at the resolution level of $K_m$. Let $\beta(Q) = \beta_Q(\gamma) = \frac{\rho(Q)}{\ell(Q)}$. Note that Jones square beta-number $\beta^2(\gamma)$ of set $K$ is $\sum_{Q \in \mathcal{Q}} \beta^2(Q)\ell(Q)$. For each term in the sum, we call $\beta^2(Q)\ell(Q)$ the local square beta-number at $Q$. We build a tour $f_m : [0,1] \to \mathbb{R}^n$ of $L_m$ by patching the tour $f_{m-1}$ locally according to the local topology of $K_m$ given by the constriction so that the sequence of tours $\{f_m\}$ converges computably.

Since the tour we build is computable, which requires parameterized approximation, the approximation scheme in computing the points in $L_m$ is not harmful.

As we mentioned earlier that points in $L_m$ may not lie in $K$, thus it is possible that, at some stage, a point chosen earlier is discovered to be outside $K$. However, when this happens, we don't remove the point. Instead, we keep such points in order to maintain the convergence of the parameterizations of the sequence of tours. Therefore, due to the inability to computably choose points strictly from $K$, we may introduce extra length to the tours. However the extra length turns out to be bounded by the local square beta-numbers and thus the access to the set $K$ in Jones's original construction is a nonessential feature of the analyst's traveling salesman problem and our characterization using Jones constriction is a proper relaxation of Jones's characterization. However, we also note that in Jones's world, using $K$ is equivalent to using the constriction.

Before getting into the construction, we describe some sub-routines that we will use in the construction to patch the tours.

First note again that, at each stage $m$, we use a precision parameter of $r = 2^m$ for points and treat them as if they have dyadic rational coordinates. It is also easy to make sure that for each $f_m$, for all $p \in [0,1]$ such that $f_m(p) \in L_m$, then $p \in [0,1] \cap \mathbb{Q}$. Thus, we may keep a table of all $p \in [0,1]$ that $f_m(p) \in L_m$.

The first procedure is $\text{attach}(f, z, x, m)$ with $z \in L_{m-1}$ or $z \in L_m$ being already explicitly traversed by $f$. This procedure modifies $f$ so that the output $f' = \text{attach}(f, z, x, m)$ traverses line segment $[z, x]$ in addition to the set $f$ originally traverses and for all $p \in [0,1]$, $f(p) - f'(p) \leq 2^{1-m}$.

The procedure first looks up the table and find $q \in [0,1]$ such that $f(q) = z$. Then it find $a \in \mathbb{Q} \cap (0,1)$ such that $|f(q - 2a) - f(q)| < 2^{1-m}$, $|f(q + 2a) - f(q)| < 2^{1-m}$, and $z$ is the only point in $L_{m-1} \cap f([q-2a, q+2a])$ and it appears only once. The output $f'$ is such that for all $p \in [0,1] \setminus [q-2a, q+2a]$, $f'(p) = f(p)$; $f'$ maps $[q-2a, q-a]$ to $f([q-2a, q])$ linearly; $f'$ maps $[q-a, q]$ to $[z, x_0]$ linearly; $f'$ maps $[q, q+a]$ to $[x_0, z]$ linearly; $f'$ maps $[q+a, q+2a]$ to $f([q, q+2a])$ linearly.

The second procedure is reconnect$(f, z_1, z_2, x_0, \ldots, x_N, m)$ with the assumption that $f$ traverses line segment $[z_1, z_2]$ from one end to the other. This procedure first looks up the table and, without loss of generality, we assume that it finds the smallest interval $[q_1, q_2] \subseteq [0,1]$ such that $f(q_1) = z_1$ and $f(q_2) = z_2$ and $f([q_1, q_2]) = [z_1, z_2]$. We obtain $f'$ by reparameterizing $f$ to include $x_0, \ldots, x_N$ in order. First we pick rational points $q_1, \ldots, q_{N-1}$ such that for each $i \in [1..N-1]$, $|f(q_i) - x_i| \leq 2\epsilon_0 3A 2^{-m}$. Then we let $f'$ map $[p, q_1]$ to $[x_0, x_1]$ and let $f'$ map $[q_{N-1}, q]$ to $[x_{N-1}, x_N]$. For $i \in [1..N-2]$, let $f'$ map $[q_i, q_{i+1}]$ to $[x_i, x_{i+1}]$. Note that if all these points involved lie in a very narrow strip, it is guaranteed that the newly added line segments are very close to the longer line segment they replace. The distance between the new parameterization and the old one is bounded by $2\epsilon_0 3A 2^{-m}$.

Note that in each of the above procedures, when $f$ is reparameterized to obtain $f'$, the table that saves the information on the preimages of points in $L_{m-1}$ and $L_m$ is updated to reflect the changes.

**Stage 0:** $m = 0$ and the size of $Q$ we consider is $\ell(Q) = A$. $L_0$ contains the two diagonal points of $[0, 1/\sqrt{n}]^n$, i.e., $L_0 = \{(0, \ldots, 0), (1/\sqrt{n}, \ldots, 1/\sqrt{n})\}$. And let $f_0$ maps $[0,1]$ linearly to the line segment $[(0, \ldots, 0), (1/\sqrt{n}, \ldots, 1/\sqrt{n})]$.

**Stage m:** For any point $z$ and $x$ with $z \neq x$, let

$$E_{z,x} = \{y \mid y - z \text{ is at most } \tfrac{2}{3}\pi \text{ from } x - z \}.$$

For all $x \in K$, let $Q_x$ be such that $x \in Q_x$ and $Q_x \in \mathcal{Q}_{m-k_0}$. Let $z_x \in L_{m-1}$ be the closest neighbor of $x$ $(2^{-m} - \sqrt{n}2^{-2^m} \leq |x_0 - z| \leq 2^{1-m} + \sqrt{n}2^{-2^{m-1}})$.

First we build a set of points that we eventually add into $L_{m-1}$ to form $L_m$. The following piece of code first find new points in $K_m$ that correspond to the cases where "farthest insertion" is required. Note that in this case, as long as the point we pick is sufficiently close to the farthest point, the construction will work. (By "sufficiently close", we mean that the point we pick is close to a farthest point enough so that another instance of "farthest insertion" does not happen within $k_0$ stages in that neighborhood.) This allows us to computably pick points for "farthest insertion" without worrying about not being able to pick the actual farthest points.

$L' \subseteq K_m$ be a set of points (dyadic cubes) such that $L_{m-1} \cup L'$ satisfies conditions C1 and C2;
$L' = L' \cap \{x \in K_m \mid \beta(Q_x) < \epsilon_0$ and $L_{m-1} \cap B_{A2^{1-m} + \sqrt{n}2^{-2^m}}(z_{x_0}) \cap E_{z_{x_0}, x_0} \setminus \{z_{x_0}\} = \varnothing\}$;
$\hat{L} = \varnothing$;
**for** all $x_0 \in L'$ **do**
   **if** $\ell([x_0, z_{x_0}]) \geq \max\{\ell([x, z_x]) \mid x \in E_{z_{x_0}, x_0} \cap B_{2^{1-m}}(z_{x_0}) \cap K_m\} - \sqrt{n}2^{-2^m}$;
   **then**
     $\hat{L} = \hat{L} \cup \{x_0\}$;
   **else**

    let $x_0' \in K_m$ be such that
        $\ell([x_0', z_{x_0}]) = \max\{\ell([x, z_{x_0}]) \mid x \in E_{z_{x_0}, x_0} \cap K_m \cap B_{2^{1-m}}(z_{x_0})\} - \sqrt{n}2^{-2^m}$;
    /* $z_{x_0'} \equiv z_{x_0}$ */
    $\hat{L} = \hat{L} \cup \{x_0'\}$;
  **end if**
**end for**

Let $\hat{L}_1 = \hat{L}$ /* $\hat{L}_1$ contains all the "farthest insertion" points */
Greedily add more points into $\hat{L}$ so that $\hat{L}$ satisfies conditions C1 and C2;

We connect every point in $\hat{L}$ to some points in $L_{m-1}$ by reparameterizing $f_{m-1}$ to get $f_m$. Initially, let $L_m = L_{m-1}$ and $f_m = f_{m-1}$. We divide the process into 3 steps.

Step 1: Farthest Insertion

    **for** all $x_0 \in \hat{L}_1$ **do** /* $\beta(Q_{x_0}) < \epsilon_0$ */
      **if** $|\hat{L} \cap E_{z_{x_0}, x_0} \cap B_{2^{1-n}}(z_{x_0}) \setminus \{x_0\}| = 0$
      **then**
        $L_m = L_m \cup \{x_0\}$;
        $f = \text{attach}(f, z_{x_0}, x_0, m)$;
      **else** /* $|\hat{L} \cap E_{z_{x_0}, x_0} \cap B_{2^{1-m}}(z_{x_0}) \setminus \{x_0\}| = 1$ */
        Let $x_1 \in \hat{L} \cap E_{z_{x_0}, x_0} \cap B_{2^{1-m}}(z_{x_0})$ with $x_1 \neq x_0$;
        $L_m = L_m \cup \{x_0, x_1\}$;
        $f = \text{attach}(f, z_{x_1}, x_1, m)$; $f = \text{attach}(f, x_1, x_0, m)$;
      **end if**
    **end for**

Step 2: Nearest Insertion

    **for** $x_0 \in \hat{L}$ with $\beta(Q_{x_0}) < \epsilon_0$ that are not processed yet **do**
      Let $z_1$ be the closest neighbor of $x_0$ in $L_{m-1} \cap B_{A2^{1-m}}(z_{x_0}) \cap E_{z_{x_0}, x_0} \setminus \{z_{x_0}\}$;
      /* Note that $f$ already explicitly traverses $[z_{x_0}, z_1]$ */
      Let $\{x_*, x_1, \ldots, x_N\} = \hat{L} \cap E_{z_{x_0}, x_0} \cap B_{\ell([z_{x_0}, z_1])}(z_{x_0})$ be ordered by $x$ component;
      **if** $x_* \neq x_0$ **then continue; end if**
      $f = \text{reconnect}(f, z_{x_0}, z_1, x_0, \ldots, x_N, m)$;
      $L_m = L_m \cup \{x_0, x_1, \ldots, x_N\}$;
      mark $x_0, x_1, \ldots, x_N$ as processed and never process again;
    **end for**

Step 3:
>   **for** all $x_0 \in \hat{L}$ with $\beta(Q_{x_0}) \geq \epsilon_0$ **do**
>     **if** $[z_{x_0}, x_0]$ is not explicitly traversed by $f$ **then** $f = \text{attach}(f, z_{x_0}, x_0, m)$;
>     **for** all $x_1 \in 3Q_{x_0} \cap (\hat{L} \cup L_{m-1})$ **do**
>       **if** $[x_0, x_1]$ is not explicitly traversed by $f$ **then** $f = \text{attach}(f, x_0, x_1, m)$;
>     **end for**
>     $L_m = L_m \cup \{x_0\}$;
>   **end for**

By construction, for every $m \in \mathbb{N}$, the distance between $f_m$ and $f_{m+1}$ is bounded by $\sqrt{n}3A2^{-m}$. So by the convergence of the geometric series, $\{f_m\}$ is a convergent sequence of bounded continuous functions. Thus $f = \lim_{m \to \infty} f_m$ exists and is actually computable, since each $f_m$ is computable from the computable constriction and the modulus of computation may be obtained by using the geometric series for the distance between $f_m$ and $f_{m+1}$.

## A.3 The Proof

In this section, we analyze the construction and prove that if Jones square beta-number of $\gamma$ is finite, then $K = \kappa(\gamma) \subseteq f([0,1])$ and $\text{length}(f) < \infty$.

*Proof.* In order to make the analysis possible, we associate with each $z \in \bigcup_{m \in \mathbb{N}} L_m$ a variable $M(z)$ and a variable $V(z)$. Variables $M$ may be taken as a savings account where local square beta-numbers are saved at times when they are not used up. The saved values are then used to cover the cost at times when new local square beta-numbers may not cover the cost. Variables $V$ are used to keep track of the information about the local environment of each point $z \in \bigcup_{m \in \mathbb{N}} L_m$ during the construction. The initial value of $M(z)$ before the first assignment is 0 and that of $V(z)$ is $\emptyset$. $M(z)$ only changes when a new assignment occurs. The values of the variables may change over stages and during the various steps of the construction in a single stage, so $M(z)$ and $V(z)$ always refer to their respective current values.

In the following, we describe how the values of variables $M$ and variables $V$ are updated during each stage and each step of the construction. We also analyze the construction and argue that, any at time during the construction, the increment to $M$ values is bounded by corresponding local square beta-numbers and $M$ values are always sufficient to cover the construction cost when local square beta-numbers may not be used. Since $M$ values come from local square beta-numbers, the increase of the length is again bounded by local square beta-numbers, though indirectly. During the construction, whenever we use $M$ values, we decrement $M$ values accordingly to ensure that $M$ values are not used repeatedly.

Since the construction is inductive, the analysis is also inductive. We will show that the following two properties hold during the construction for all $z \in L_m$, $m \in \mathbb{N}$.

**P1:** For all $z' \in V(z)$, let $\{y_1, \ldots, y_N\} = V(z)$ be arranged in the order of their projections on the line determined by $[z, z']$. Then for all $j \leq N-1$, $[y_j, y_{j+1}]$ is a direct line segment in $f_m$.

**P2:** $V(z) \neq \emptyset$ and one of the following is true.

(1) If there are at least two points $z_1, z_2 \in V(z)$ such that the angle between $[z, z_1]$ and $[z, z_2]$ is at least $2\pi/3$, then $M(z) \geq \sum_{z' \in V(z)} \ell([z, z'])$.

(2) If for some $z' \neq z$, $E_{z,z'} \cap V(z) = \emptyset$ and $V(z) \neq \emptyset$, then we have both of the following.

   (a) $M(z) \geq 2^{1-m} + \sum_{z' \in V(z)} \ell([z, z'])$.
   (b) For all $k \geq 0$, if $B_{2^{-m-k}}(z) \cap E_{z,z'} \neq B_{2^{1-m}}(z) \cap E_{z,z'}$ (at the resolution of $K_m$), then $M(z) \geq A 2^{1-m-k} + \sum_{z' \in V(z)} \ell([z, z'])$.

We verify that the properties are true initially and that if the properties are true at any time, after any legal step of construction the properties are still true.

**Stage 0:** Initially, $M$ values are all $0$ and $V$ values are all $\emptyset$, so the properties trivially hold.

Let the two diagonal points be $z_1, z_2$. Note that $\ell([z_1, z_2]) = 1$. Let $M(z_1) = A + 1$ and $M(z_2) = A + 1$. Let $V(z_1) = \{v_2\}$ and $V(z_2) = \{v_1\}$. Note that this assignment may be regarded as a special case for step 3 in the construction. Without loss of generality, assume $z_1$ is added before $z_2$. It is easy to check that property P1 and property P2 (part (2)) are true after $z_1$ is added and remain true when $z_2$ is added.

**Stage m:** We give different assignment rules for $M$ values for each of the 3 steps in the construction. For clarity, we keep the code for the construction and give the assignment rules in annotations.

Step 1: Farthest Insertion

```
for all x_0 ∈ L̂_1 do /* β(Q_{x_0}) < ε_0 */
   if |L̂ ∩ E_{z_{x_0}, x_0} ∩ B_{2^{1-m}}(z_{x_0})| = 1
   then
      L_m = L_m ∪ {x_0};
      f = attach(f, z_{x_0}, x_0, m);
      @ V(x_0) = V(x_0) ∪ {z_{x_0}};
      @ if V(z_{x_0}) ∩ E_{z_{x_0}, x_0} ≠ ∅
      @ then
      @    V(z_{x_0}) = V(z_{x_0}) \ V(z_{x_0}) ∩ E_{z_{x_0}, x_0};
      @ end if
      @ V(z_{x_0}) = V(z_{x_0}) ∪ {x_0};
      @ M(z_{x_0}) = M(z_{x_0}) − A 2^{1-m} + 2^{1-m};
      @ M(x_0) = 2 · 2^{1-m};
   else /* |L̂ ∩ E_{z_{x_0}, x_0} ∩ B_{2^{1-m}}(z_{x_0}) \ {x_0}| = 1 */
      Let x_1 ∈ L̂ ∩ E_{z_{x_0}, x_0} ∩ B_{2^{1-m}}(z_{x_0}) with x_1 ≠ x_0;
      L_m = L_m ∪ {x_0, x_1};
      f = attach(f, z_{x_1}, x_1, m); f = attach(f, x_1, x_0, m);
```

```
    @ V(x_0) = V(x_0) ∪ {x_1};
    @ V(x_1) = V(x_1) ∪ {z_{x_0}, x_0};
    @ if V(z_{x_0}) ∩ E_{z_{x_0},x_0} ≠ ∅
    @ then
    @     V(z_{x_0}) = V(z_{x_0}) \ V(z_{x_0}) ∩ E_{z_{x_0},x_0};
    @ end if
    @ V(z_{x_0}) = V(z_{x_0}) ∪ {x_1};
    @ M(z_{x_0}) = M(z_{x_0}) − A2^{1−m} + 2^{1−m} + 2√n 2^{−2^{m−1}};
    @ M(x_0) = 2(2^{1−m} + 2√n 2^{−2^{m−1}});
    @ M(x_1) = 2(2^{1−m} + 2√n 2^{−2^{m−1}});
  end if
end for
```

Whenever "farthest insertion" is involved, the point $x_0$ under consideration always lies in a narrow cube that contains $x_0$, $z_{x_0}$, and possibly $x_1$. Therefore, P1 is satisfied at $x_0$ due to the narrowness of the cube. For $z_{x_0}$, P1 is maintained due to the removal of points in $V(z_{x_0}) \cap E_{z_{x_0},x_0}$ from $V(z_{x_0})$.

In every stage $m \in \mathbb{N}$, the tour $f_m$ traverses a set of line segments. By the construction, every line segment is traversed at most twice. Therefore, for each $m \in \mathbb{N}$, length$(f_m) \leq 2\ell(f_m([0,1]))$, where $\ell(f_m([0,1]))$ is the one dimensional Hausdorff measure of the set $(f_m([0,1]))$. In the following analysis, we bound $\ell(f_m([0,1]))$ instead of length$(f_m)$.

The length of each line segment we add in this case is at most $2^{1-m} + 2\sqrt{n}2^{-2^{m-1}}$ (taking into consideration of the approximation of the locations of end points), and we add at most 2 line segments. The total $M$ values for $z$, $x_0$, and $x_1$ (if it exists) is bounded by $5(2^{1-m} + 2\sqrt{n}2^{-2^{m-1}})$. So the sum of added length and $M$ values is bounded by $7 \cdot 2^{1-m}$.

Since $A > 9$, it suffices to show that we may use $A2^{1-m}$ from old $M$ value to cover the cost here.

Before this step of construction involving $x_0$ and $z_{x_0}$, $z_{x_0}$ satisfied property P2.

If part (1) of property P2 was satisfied before this step, there is a point $z' \in V(z_{x_0}) \cap E_{z_{x_0},x_0}$ such that $\ell([z_{x_0}, z']) > A2^{1-m}$. Since $z'$ is removed from $V(z_{x_0})$, the reduction of $A1^{1-m}$ from $M(z_{x_0})$ is used to cover the cost and is balanced by the removal of $z'$.

If after the addition of either $x_0$ or $x_1$ to $V(z_{x_0})$, the condition of part (1) in property P2 is true, then since the addition to $M(z_{x_0})$, which is $2^{1-m} + 2\sqrt{n}2^{-2^{m-1}} \geq \ell([z_{x_0}, x_0])$ (or in case $|\hat{L}_1 \cap E_{z_{x_0},x_0} \cap B_{2^{1-m}}(z_{x_0}) \setminus \{x_0\}| = 1$, $2^{1-m} + 2\sqrt{n}2^{-2^{m-1}} \geq \ell([z_{x_0}, x_1])$), part (1) in property P2 remains true.

If after the addition of either $x_0$ or $x_1$ to $V(z_{x_0})$, the condition of part (2) in property P2 is true, then since the addition to $M(z_{x_0})$ is $2^{1-m} + 2\sqrt{n}2^{-2^{m-1}}$, part (2)-(a) in property P2 is satisfied at $z_{x_0}$. Since $\beta(Q_{x_0}) < \epsilon_0$, on the side of $z_{x_0}$ (given by $z'$ in the P2) where $V(z_{x_0}) \cap E_{x_{x_0},z'}$ is empty, there will not be further construction within less than $k_0$ stages, i.e., the condition of part (2)-(b) of property P2 will not be true within $k_0$ stages. Together with the fact that $2^{1-m} \geq A2^{1-m-k_0}$, part (2)-(b) of property P2 is satisfied at $z_{x_0}$.

$V(x_0)$ contains only one point whose distance from $x_0$ is between $2^{-m} - 2^{-2^{m-1}}$ and $2^{1-m} + 2^{-2^{m-1}}$. So part (2)-(a) of property P2 is satisfied at $x_0$. Since $\beta(Q_{x_0}) < \epsilon_0$, there will be no further construction within less than $k_0$ stages on the empty side of $V(x_0)$, i.e., the condition of part (2)-(b) of property P2 will not be true within $k_0$ stages. Therefore, part (2)-(b) of property P2 is satisfied at $x_0$.

If $x_1$ is added to $L_m$ in this step, since $\beta(Q_{x_0}) < \epsilon_0$, $x_1$ is between $z_{x_0}$ and $x_0$, part (1) of property P2 is satisfied at $x_1$.

If part (2) was satisfied before this step, we have two possibilities.

One possibility is that $E_{z_{x_0},x_0} \cap V(z_{x_0}) = \emptyset$. Then since we have a "farthest insertion" construction at $x_0$, $B_{2^{-m}}(z_{x_0}) \cap E_{z_{x_0},x_0} \neq B_{2^{1-m}}(z_{x_0}) \cap E_{z_{x_0},x_0}$, i.e., the condition for part (2)-(b) of property P2 is true and thus $M(z_{x_0}) \geq A2^{1-m} + \sum_{z' \in V(z_{x_0})} \ell([z_{x_0}, z'])$. Now the extra $A2^{1-m}$ may be used to cover the cost and is the amount that is deducted from $M(z_{x_0})$. After we add $x_0$ to $V(z_{x_0})$, since $\beta(Q_{x_0}) < \epsilon_0$, the condition of part (1) of property P2 is true. Since $2^{1-m} + 2\sqrt{n}2^{-2^{m-1}} \geq \ell([z_{x_0}, x_0])$ (or in case $|\hat{L} \cap E_{z_{x_0},x_0} \cap B_{2^{1-m}}(z_{x_0}) \setminus \{x_0\}| = 1$, $2^{1-m} + 2\sqrt{n}2^{-2^{m-1}} \geq \ell([z_{x_0}, x_1])$), part (1) of property P2 is satisfied at $z_{x_0}$.

The other possibility is that $E_{z_{x_0},x_0} \cap V(z_{x_0}) \neq \emptyset$. Then there is a point $z' \in V(z_{x_0}) \cap E_{z_{x_0},x_0}$ such that $\ell([z_{x_0}, z']) > A2^{1-m}$. Now the analysis will be the same as in the case when part (1) of property P2 was satisfied before this step except that we need to note that although $V(z_{x_0})$ changes, the amount $M(z_{x_0}) - \sum_{z' \in V(z_{x_0})} \ell([z_{x_0}, z'])$ does not decrease during the process. Therefore part (2) of property P2 remains true and thus P2 remains true.

The analysis of the properties at $x_0$ and $x_1$ are the same as in the case in the case when part (1) of property P2 was satisfied before this step.

Also note that we never make variable $V$ empty.

Step 2: Nearest Insertion

    **for** all $x_0 \in \hat{L}$ with $\beta(Q_{x_0}) < \epsilon_0$ that are not processed yet **do**
        Let $z_1$ be the closest neighbor of $x_0$ in $L_{m-1} \cap B_{A2^{1-m}}(z_{x_0}) \cap E_{z_{x_0},x_0} \setminus \{z_{x_0}\}$;
        /* Note that $[z_{x_0}, z_1]$ is traversed explicitly by $f_{m-1}$ */
        Let $\{x_*, x_1, \ldots, x_N\} = \hat{L} \cap E_{z_{x_0},x_0} \cap B_{\ell([z_{x_0},z_1])}(z_{x_0})$ be ordered by $x$ component;
        **if** $x_* \neq x_0$ **then continue; end if**
        $f = \text{reconnect}(f, z_{x_0}, z_1, x_0, \ldots, x_N, m)$;
        @ $V(z_{x_0}) = V(z_{x_0}) \cup \{x_0\} \setminus \{z_1\}$;

@ $M(z_{x_0}) = M(z_{x_0}) - \ell([z_{x_0}, z_1]) + \ell([z_{x_0}, x_0])$;
@ $V(x_0) = V(x_0) \cup \{z_{x_0}\}$;
@ $M(x_0) = M(x_0) + \ell([z_{x_0}, x_0])$;
@ $V(z_1) = V(z_1) \cup \{x_N\} \setminus \{z_{x_0}\}$;
@ $M(z_1) = M(z_1) - \ell([z_{x_0}, z_1]) + \ell([x_N, z_1])$;
@ $V(x_N) = V(x_N) \cup \{z_1\}$;
@ $M(x_N) = M(x_N) + \ell([x_N, z_1])$;
**for** $i = 0$ **to** $N - 1$ **do**
   @ $V(x_i) = V(x_i) \cup \{x_{i+1}\}$;
   @ $M(x_i) = M(x_i) + \ell([x_i, x_{i+1}])$;
   @ $V(x_{i+1}) = V(x_{i+1}) \cup \{x_i\}$;
   @ $M(x_{i+1}) = M(x_{i+1}) + \ell([x_i, x_{i+1}])$;
**end for**
$L_m = L_m \cup \{x_0, x_1, \ldots, x_N\}$;
mark $x_0, x_1, \ldots, x_N$ as processed and never process again;
**end for**

Since in this case, the points we work with are all located along a very narrow and long cylinder, by Pythagorean, we have that the length added is bounded by

$$C_3 \sum_{\beta(Q) < \epsilon_0} \beta(Q)^2 \ell(Q).$$

Note that if we make $\epsilon_0$ smaller, constant $C_3$ can also be chosen smaller. Since we don't need to increase $C_3$, we may fix $C_3$ large enough for all sufficiently small $\epsilon_0$ so that $C_3$ does not depend on the choice of $\epsilon_0$ or the choice of $A$. Also since the changes happens in a narrow cylinder, P1 is maintained.

For $j \in [0..N]$, $M(x_j)$ satisfies P2, in particular part (1) of P2, since each of them is connected to 2 other points that are more than $2\pi/3$ angle apart.

For $z_{x_0}$, in this case, $z_1 \in V(z_{x_0})$ before we make the changes. So $E_{z_{x_0}, x_0} \cap V(z_{x_0}) \neq \emptyset$, and after we make the changes to $M(z_{x_0})$, since $V(z_{x_0})$ is changed accordingly, the value $M(z_{x_0}) - \sum_{z' \in V(z_{x_0})} \ell([z_{x_0}, z'])$ does not decrease. Therefore P2 remains true after this step regardless of whether part (1) or part (2) was true. The same argument tells us that P2 remains true at $z_1$.

Due to the way we assign $M$ values, the total increment of $M$ values in this case is bounded by at most 2 times the total increase of length, i.e.,

$$2 \cdot C_3 \sum_{\beta(Q) < \epsilon_0} \beta(Q)^2 \ell(Q).$$

Step 3:
   **for** all $x_0 \in \hat{L}$ with $\beta(Q_{x_0}) \geq \epsilon_0$ **do**

    **if** $[z_{x_0}, x_0]$ is not explicitly traversed by $f$ **then**
        $f = \text{attach}(f, z_{x_0}, x_0, m)$;
        @ $V(x_0) = V(x_0) \cup \{z_{x_0}\}$;
        @ $M(x_0) = M(x_0) + \ell([x_0, z_{x_0}])$;
        @ $V(z_{x_0}) = V(z_{x_0}) \cup \{x_0\}$;
        @ $M(z_{x_0}) = M(z_{x_0}) + \ell([x_0, z_{x_0}])$;
    **end if**
    **for** all $x_1 \in 3Q_{x_0}(\hat{L} \cup L_{m-1})$ **do**
        **if** $[x_0, x_1]$ is not explicitly traversed by $f$ **then**
            $f = \text{attach}(f, x_0, x_1, m)$;
            @ $V(x_0) = V(x_0) \cup \{x_1\}$;
            @ $M(x_0) = M(x_0) + \ell([x_0, x_1])$;
            @ $V(x_1) = V(x_1) \cup \{x_0\}$;
            @ $M(x_1) = M(x_1) + \ell([x_0, x_1])$;
        **end if**
    **end for**
    $L_m = L_m \cup \{x_0\}$;
    @ $M(x_0) = M(x_0) + A2^{-m}$;
**end for**

It is easy to verify that property P1 is maintained for each involved point.

Since we assign $A2^{-m}$ to $M(x_0)$ in addition to the sum of length of connected line segments, P2 is true for every $x_0$. For those $x_1 \in L_{m-1}$ that are involved in this case, $M(x_1)$ value is incremented by the length of the line segment for each of the added line segment. The value $M(x_1) - \sum_{z' \in V(x_1)} \ell([x_1, z'])$ does not decrease. Therefore, P2 remains true after the changes.

Let $C_1$ be the maximum number of points that can be fit into $3Q$ and satisfy property C1. Let $C_2$ be the maximum number of points in $L_m \setminus L_{m-1}$ that can fit into $3Q$. Note that $C_1$ and $C_2$ are functions of $n$, which is the dimension of the Euclidean space we are working with. So both the total length we add to $f_m$ and for each point in $L_m$, the total increment of $M$ value are bounded by

$$C_1 \cdot A2^{-m} + C_1 \cdot 2 \sum_{\beta(Q) \geq \epsilon_0} C_2 \cdot 3\sqrt{n}\ell(Q) = \frac{9 \cdot C_1 \cdot C_2 \sqrt{n}}{\epsilon_0^2} \sum_{\beta(Q) \geq \epsilon_0} \epsilon_0^2 \ell(Q)$$

$$\leq \frac{9 \cdot C_1 \cdot C_2 \sqrt{n}}{\epsilon_0^2} \sum_{\beta(Q) \geq \epsilon_0} \beta(Q)^2 \ell(Q).$$

We have, by now, established case by case bound on length increment in every stage. Now we put all these things together and bound the length of the tour we obtain.

Let
$$M_m = \sum_{z \in L_m} M(z),$$
where $M(z)$ takes the value at the end of stage $m$. So $M_0 = 2A + 2$.

Let $l_m$ be the total increment of length from $f_{m-1}$ to $f_m$ introduced by "farthest insertion" and $l_0 = 0$.

Let $C = \max\left(\frac{9 \cdot C_1 \cdot C_2 \sqrt{n}}{\epsilon_0^2}, 2 \cdot C_3\right)$.

Let $M_{m,1}$ be the total reduction of $M$ values in stage $m$ in "farthest insertion". Let $M_{m,23}$ be the total increment of $M$ values in stage $m$ in Steps 2 and 3. By the construction, $M_{m,23} \leq C \sum_{Q \in \mathcal{Q}_{m-k_0}} \beta(Q)^2 \ell(Q)$.

Note that in an instance of "farthest insertion", the increment of length $\Delta l$ is bounded by $2(2^{1-m} + 2\sqrt{n}2^{-2^{m-1}})$, i.e., $\Delta l \leq 2(2^{1-m} + 2\sqrt{n}2^{-2^{m-1}}) \leq 3 \cdot 2^{1-m}$. For the involved point $z \in L_{m-1} \subset L_m$ and $x_0, x_1 \in L_m \setminus L_{m-1}$, the increment of $M$ values at $z$, $x_0$, and $x_1$ is at most by $5(2^{1-m} + 2\sqrt{n}2^{-2^{m-1}}) \leq 7 \cdot 2^{1-m}$ and the loss of $M$ value at $z$ is $A2^{1-m}$. Note that $x_1$ may not be present in the construction. Since we give an upper bound here, we use the worst case and assume $x_1$ is present. So the total reduction in $M$ value involved in such an instance of "farthest insertion", $\Delta M(z)$ is at least $(A - 5)2^{-m+1}$. So for each individual instance of "farthest insertion" in stage $m$, the ratio between the reduction in $M$ values and the increment of length is
$$\frac{\Delta M(z)}{\Delta l} \geq \frac{A - 7}{3}.$$

So $M_{m,1} \geq \frac{A-7}{3} l_m$.

Note that in the following, we are combining the $\beta(Q) \geq \epsilon_0$ part and the $\beta(Q) < \epsilon_0$ part of the sum of local square beta-numbers, i.e., the sums for Step 2 and Step 3 are combined.

$$M_m - M_{m-1} = M_{m,23} - M_{m,1} < C \sum_{Q \in \mathcal{Q}_{m-k_0}} \beta(Q)^2 \ell(Q) - \frac{A-7}{3} l_m.$$

Note that due to property P2, for all $m_0 \in \mathbb{N}$, $M_{m_0} \geq 0$. So

$$0 \leq M_{m_0} = M_0 + \sum_{m=1}^{m_0} (M_m - M_{m-1}) < M_0 + \sum_{m=1}^{m_0} \left( C \sum_{Q \in \mathcal{Q}_{m-k_0}} \beta(Q)^2 \ell(Q) - \frac{A-7}{3} l_m \right).$$

Therefore
$$\sum_{m=1}^{m_0} \frac{A-7}{3} l_m < M_0 + \sum_{m=1}^{m_0} \left( C \sum_{Q \in \mathcal{Q}_{m-k_0}} \beta(Q)^2 \ell(Q) \right).$$

And thus
$$\sum_{m=1}^{\infty} \frac{A-7}{3} l_m \leq M_0 + C \sum_{m=1}^{\infty} \left( \sum_{Q \in \mathcal{Q}_{m-k_0}} \beta(Q)^2 \ell(Q) \right).$$

So
$$\sum_{m=1}^{\infty} l_m \leq \frac{3M_0}{A-7} + \frac{3C}{A-7} \sum_{m=1}^{\infty} \sum_{Q \in \mathcal{Q}_{m-k_0}} \beta(Q)^2 \ell(Q).$$

By our construction, $\ell(f_m) - \ell(f_{m-1})$ consists of the increments in Step 1, Step 2, and Step 3. So
$$\ell(f_m) - \ell(f_{m-1}) \leq l_m + C \sum_{Q \in \mathcal{Q}_{m-k_0}} \beta(Q)^2 \ell(Q).$$

Now we have that the one dimensional Hausdorff measure of $f([0,1])$ is

$$\lim_{m \to \infty} \ell(f_m) = \ell(f_0) + \sum_{m=1}^{\infty} (\ell(f_m) - \ell(f_{m-1}))$$
$$\leq \ell(f_0) + \sum_{m=1}^{\infty} \left( l_m + C \sum_{Q \in \mathcal{Q}_{m-k_0}} \beta(Q)^2 \ell(Q) \right)$$
$$= \ell(f_0) + C \sum_{m=1}^{\infty} \sum_{Q \in \mathcal{Q}_{m-k_0}} \beta(Q)^2 \ell(Q) + \sum_{m=1}^{\infty} l_m$$
$$\leq \ell(f_0) + C \sum_{m=1}^{\infty} \sum_{Q \in \mathcal{Q}_{m-k_0}} \beta(Q)^2 \ell(Q) + \frac{3M_0}{A-7} + \frac{3C}{A-7} \sum_{m=1}^{\infty} \sum_{Q \in \mathcal{Q}_{m-k_0}} \beta(Q)^2 \ell(Q)$$
$$= \ell(f_0) + \frac{3M_0}{A-7} + C\left(1 + \frac{3}{A-7}\right) \sum_{m=1}^{\infty} \sum_{Q \in \mathcal{Q}_{m-k_0}} \beta(Q)^2 \ell(Q).$$

Therefore
$$\text{length}(f) \leq 2 \cdot \mathcal{H}^1(f([0,1])) \leq 2\ell(f_0) + \frac{6M_0}{A-7} + 2C\left(1 + \frac{3}{A-7}\right) \sum_{m=1}^{\infty} \sum_{Q \in \mathcal{Q}_{m-k_0}} \beta(Q)^2 \ell(Q).$$

Since the square beta-number $\beta^2(\gamma) < \infty$, $\text{length}(f) < \infty$. $\square$